\documentclass{ws-p8-50x6-00a}

\begin{document}

\noindent
\begin{minipage}[t]{.2\linewidth}
\leavevmode
 \hspace*{-.8cm}
\psfig{file=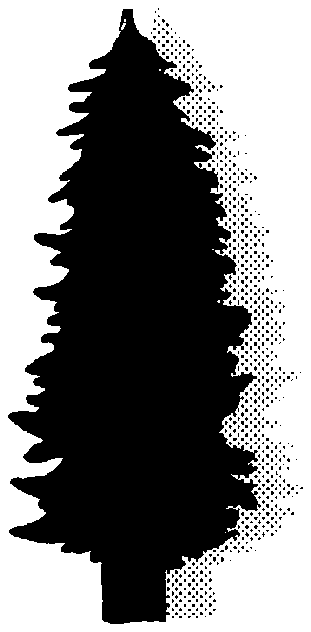,width=3cm}
\end{minipage} \hfill
\begin{minipage}[b]{.45\linewidth}
\rightline{SCIPP 99/39}
\rightline{September 1999}
\vspace{3cm}
\end{minipage}
\vskip1.5cm

\thispagestyle{empty}

\begin{center}
{\large\bf PRECISE ELECTROWEAK MEASUREMENTS AT THE $Z^0$ POLE}\\[2pc]

BRUCE A. SCHUMM \\

Institute for Particle Physics \\
University of California, Santa Cruz
\end{center}
\vskip1cm

Over the last decade, precise LEP and SLC
measurements of electroweak coupling parameters at the
$Z^0$ pole have lead to tests of the Standard Model to
unprecedented precision. This report presents a comprehensive
review of these studies, including a review of relevant $Z^0$ pole
physics issues, facilities, instrumentation, and the measurements made.
Global fits for the Higgs Boson mass and $Z^0-b$ coupling
parameters are also presented.

\vfill
\begin{center}
{\small Talk presented at the 1999 Conference on Physics In Collision,
Anne Arbor, Michigan, June 24--26, 1999}
\end{center}
\vfill
\clearpage
\setcounter{page}{1}

\title{Precise Electroweak Measurements at the $Z^0$ Pole}

\author{B. A. Schumm}

\address{University of California, Santa Cruz, CA 95064,
USA\\E-mail: schumm@scipp.ucsc.edu}


\maketitle

\abstracts{
Over the last decade,
precise LEP and SLC
measurements of electroweak coupling parameters at the
$Z^0$ pole have lead to tests of the Standard Model to
unprecedented precision. This report presents a comprehensive
review of these studies, including a review of relevant $Z^0$ pole
physics issues, facilities, instrumentation, and the measurements made.
Global fits for the Higgs Boson mass and $Z^0-b$ coupling
parameters are also presented.}

\section{Introduction}
With the end of the SLC/SLD run at SLAC in June, 1998, the era
of precision tests of the standard model using $e^+e^-$ beams tuned
to the energy of the $Z^0$-boson resonance came to a close, at least
for the foreseeable future. Within the decade since the first production
in 1988 of the $Z^0$ with lepton beams, however, the precision
of the resulting
constraints on the phenomonolgy of the Standard Model
has exceeded even the more optimistic estimates
put forth at the beginning of
the running. While no clear disagreements with the Standard Model have
come to light during this period, the development of the Standard Model
and its subsequent confirmation at the 1-loop level, to accuracies
exceeding one part in $10^{-3}$, surely stands as one of the great
achievements of human scientific pursuit.

\section{Standard Model Background}
Precise tests of the Standard Model at the $Z^0$ pole are performed
via measurements of the couplings of the various fermions to the
$Z^0$, and primarily through the quantitative extent of
parity violation in those couplings. In the Standard Model, the
neutral components of the parity-violating SU(2)$_L$ and
partially parity-conserving U(1) interactions are mixed
according to an angle $\theta_W$, the
`weak mixing angle'. The resulting physical states are
the weak $Z^0$ boson and the electromagnetic $\gamma$,
the latter of which mediates a purely parity conserving interaction,
while the $Z^0$ thus mediates a partially parity-violating interaction.
The quantitative extent of the $Z^0$ parity violation thus
depends on the value of $\theta_W$, as well as the electric ($Q$) and
weak isospin ($t_3$) charges of
the interacting fermion. Specifically, in terms of the left- and
right-handed $Z^0$-fermion couplings $g_L^f$ and $g_R^f$, the
quantitative extent of parity violation is expressed via the
parity-violation parameters $A_f$:
\begin{equation}
\label{eq:pvp}
A_f = {(g_L^f)^2 - (g_R^f)^2 \over
               (g_L^f)^2 + (g_R^f)^2 }
\end{equation}
where
$ g_L^f = t_3^f - Q^f \sin^2 \theta_W $ and
$ g_R^f =       - Q^f \sin^2 \theta_W.$
Table~\ref{tab:smp} 
shows the electroweak charges, couplings and parity violation
parameters $A_f$ for the various fermion species. Also shown is
the resulting sensitivity to the weak mixing angle.
It is seen that the extent of parity violation in the coupling
of the $Z^0$ to charged leptons is very sensitive to the value
of $\sin^2 \theta_W$, and thus to the vacuum polarization
effects which renormalize the boson propagator. On the other
hand, the extent of parity violation to down-type quarks is
quite insensitive to $\sin^2 \theta_W$, and thus
measurements of parity violation in these couplings are interesting
in that they constrain $Z^0-f$ vertex corrections independently of
the propagator effects which renormalize
$\sin^2 \theta_W$. 

\begin{table}[t]
\caption{Standard Model Couplings and Parity Violation (assuming
  $\sin^2 \theta_W = 0.231$) \label{tab:smp}}
\begin{center}
\begin{tabular}{|l|c|c|c|c|c|c|}
\hline
Fermion  &  $t_3^f$  & $Q^f$  & $g_L^f$ & $g_R^f$ & $A_f$  &
 ${\rm d} A_f / {\rm d} \sin^2 \theta_W $  \\
\hline
\hline
$e, \mu, \tau$ & -1/2 & -1   & -0.269  &  0.231  &  0.151  &  -7.8 \\
$\nu         $ & +1/2 &  0   &  0.500  &  0.000  &  1.000  &  -0.0 \\
$d,s,b       $ & -1/2 & -1/3 & -0.423  &  0.077  &  0.935  &  -0.6 \\
$u,c,t       $ & +1/2 & +2/3 &  0.346  & -0.154  &  0.669  &  -3.5 \\
\hline
\end{tabular}
\end{center}
\end{table}

\begin{figure}[t]
\epsfxsize=23pc 
\epsfbox{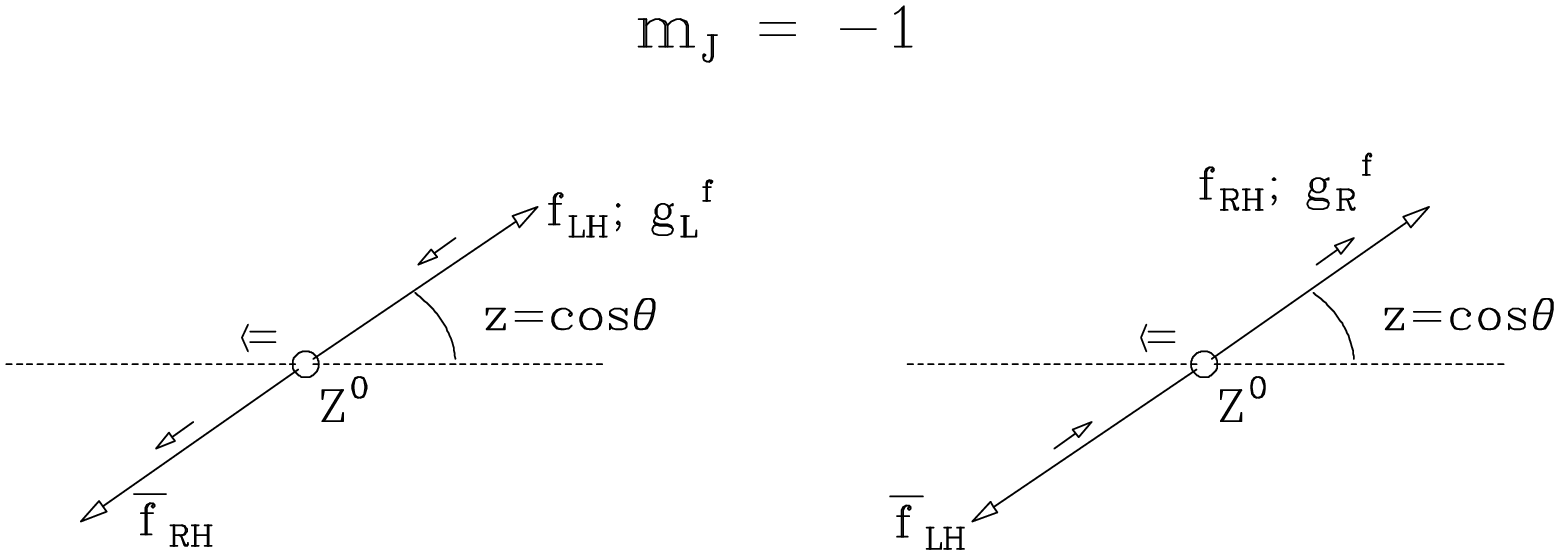} 
\caption{Decay of an $m_J = -1$ $Z^0$ boson into left- and right-
handed fermions. \label{fig:angmom}}
\end{figure}

The angular differential cross-section for fermion production at the
$Z^0$ pole is given by
\begin{equation}
\label{eq:dxs}
{d\sigma^f \over dz}
 \propto (1 - A_e P_e)(1 + z^2) +
2 A_f (A_e - P_e) z ,\nonumber
\end{equation}
where
$P_e$ is the longitudinal polarization of the electron beam, and
$z = \cos \theta$.
This relation can be understood entirely from
angular momentum arguments. Figure~\ref{fig:angmom}
shows that the angular cross section for an $m_J = -1$
$Z^0$ boson decaying to a LH fermion and RH antifermion will be
proportional to square of the spin-1 rotation matrix element
corresponding to the transformation between an $m_J = -1$ state
with spin projection on the $z$ axis and an $m_J = -1$ state with
spin projection along an axis rotated by $\cos \theta$ relative
to the $z$ axis:~\cite{BandJ}
\begin{equation}
\label{eq:ad0}
\sigma^{m_J=-1}_{LH} \propto  |d^{(1)}_{-1,-1}(z)|^2 = (1+z)^2.
\end{equation}
For decay into a RH fermion,
the argument is identical up to the substitution of $m_J = +1$ for
the spin projection onto the rotated axis:
\begin{equation}
\label{eq:ad1}
\sigma^{m_J=-1}_{RH} \propto  |d^{(1)}_{+1,-1}(z)|^2 = (1-z)^2.
\end{equation}
Since the LH and RH coupling strengths are just $g_L$ and $g_R$,
we have for an $m_J=-1$ $Z^0$ sample:
\begin{equation}
\label{eq:ad2}
\sigma^{m_J=-1}_{tot}(z) \propto g_L^2 (1 + z)^2 +
  g_R^2 (1-z)^2 = (g_L^2 + g_R^2)(1 + z^2) + 2(g_L^2 - g_R^2) z.
\end{equation}
The presence of the term odd in $z$ leads to a forward-backward
asymmetry in the overall fermion production rate, provided that
$|g_L| \neq |g_R|$, i.e., that parity is violated.
With a purely left-handed electron beam $(P_e = -1)$,
the sample of produced
$Z^0$ bosons would be purely $m_J=-1$, and the forward-backward
asymmetry would be
\begin{equation}
\label{eq:ad3}
{\rm A}_{FB}^{{m_J=-1}}(z) =
{\sigma^f({z}) - \sigma^f({-z}) \over
 \sigma^f({z}) + \sigma^f({-z}) } =
{{g_L^2} - {g_R^2} \over  {g_L^2} + {g_R^2}}
  {2z \over 1 + z^2} = {A_f}{2z \over 1 + z^2}
\end{equation}
where $A_f$ is the parity violation parameter for fermion species
$f$ introduced above. However, by reproducing the above arguments
for the case of $m_J=+1$, it's easy to see that the sign of
the asymmetry flips. Thus, a forward-backward asymmetry develops
only if there are {\it different numbers} of $m_J=+1$ and
$m_J=-1$ $Z^0$'s in the overall sample. For an unpolarized electron
beam (equal numbers of LH and RH electrons), this will only be
true to the extent that the LH and RH couplings of the {\it electron}
differ. Thus, the unpolarized forward-backward asymmetry also
depends upon the extent $A_e$ of parity violation in the
$Z^0$-$e$ coupling:
\begin{equation}
\label{eq:ad4}
{\rm A}^f_{FB}(z) =
{\sigma^f({z}) - \sigma^f({-z}) \over
 \sigma^f({z}) + \sigma^f({-z}) } =
{A_e}{A_f}{2z \over 1 + z^2},
\end{equation}
in agreement with Eqn.~(\ref{eq:dxs}).
By convention, this observable is usually expressed in terms of
the integrated forward ($z>0$) and backward ($z<0$) cross sections:
\begin{equation}
\label{eq:ad5}
{\rm A}_{FB}^f =
{\sigma_F^f - \sigma_B^f \over \sigma_F^f + \sigma_B^f} =
{3  \over 4} {A_e}{A_f}.
\end{equation}
where $\sigma^f_F = \int_0^1 \sigma^f(z)dz$
and   $\sigma^f_B = \int_0^1 \sigma^f(-z)dz$.

\subsection{Polarized Electron Beams}

With a polarized electron beam,
parity violation in the $Z^0$-$e$ coupling can be measured
directly by forming
the asymmetry $A_{LR}$ in the left- and right-handed $Z^0$ production
cross sections:
\begin{equation}
\label{eq:ad6}
 {\rm A}_{{\rm LR}} = {\sigma_{e_L^-e^+} - \sigma_{e_R^-e^+}
              \over      \sigma_{e_L^-e^+} + \sigma_{e_R^-e^+}}
=  {(g_L^e)^2 - (g_R^e)^2 \over
               (g_L^e)^2 + (g_R^e)^2 }    = A_e.
\end{equation}
The sensitivity of this parameter to $\sin^2 \theta_W$ (see
Table~\ref{tab:smp}), combined with the indifference of the quantity
to the particular final state fermion,
render this the
single most accurate approach to the measurement of $\sin^2 \theta_W$.

It is also possible to form a polarized {\it final-state} asymmetry
\begin{equation}
\label{eq:afbp}
 {\tilde {\rm A}}^f_{FB}(z) =
{[\sigma_L^f(z) - \sigma_L^f(-z)] -
 [\sigma_R^f(z) - \sigma_R^f(-z)] \over
  \sigma_L^f(z) + \sigma_L^f(-z)  +
  \sigma_R^f(z) + \sigma_R^f(-z)}
 = {A_f} {2z \over 1 + z^2}.
\end{equation}
Unlike its unpolarized counterpart, this observable is sensitive
to {\it final state} parity violation alone. In the case $f=b$,
this provides a direct constraint on $Z^0$-$b$ vertex
effects, independent of the propagator effects which alter the
value of $\sin^2 \theta_W$ (see Table~\ref{tab:smp}).

Both of these observables, however, are mitigated by incomplete
beam polarization ($|P_e| \neq 1$) according to
$$ {\rm A}_{meas} = |P_e| {\rm A}_{true}.$$
Thus, in order to measure these observables accurately, it is
necessary to have an electron beam with a large and well-measured
polarization.

\subsection{$Z^0$ Lineshape Parameters}

At Born level, the total cross section for the production
of fermion type $f$ is given by
\begin{equation}
\label{eq:ad8}
   \sigma_{e^+e^- \rightarrow f {\overline f}}(s) =
 {12 \pi \over {M_Z^2}}
 {s {\Gamma_{e^+e^-} \Gamma_{f {\overline f}}} \over
 (s - {M_Z^2})^2 + s^2 {\Gamma_Z^2}/{M_Z^2}}
\end{equation}
where $s$ is the square of the cms energy, and the partial width
$\Gamma_{f {\overline f}}$ is given by
\begin{equation}
\label{eq:pwd}
   \Gamma_{f {\overline f}} = 2 c_f
[({g_L^f})^2 + ({g_R^f})^2]
{ G_F {M_Z^3} \over 12 \sqrt{2} \pi}.
\end{equation}
Thus, measurements of the $Z^0$ resonance parameters $M_Z$,
$\Gamma_Z$, and the peak cross section
$\sigma^0$, are important for constraining the
Standard Model. In particular, the relation
\begin{equation}
  {\sin 2 \theta_W} = [{ 4 \pi \alpha_{\rm em} \over
   \sqrt{2} G_F {M_Z^2}}]^{1/2}
\label{eq:bd1}
\end{equation}
exhibits the value of a precise measurement of $M_Z$, given
our precise knowledge of $\alpha_{\rm em}$ and $G_F$. With the value
of $\theta_W$ thus constrained, measurements of
$\sin^2 \theta_W$ via parity violation become exacting
consistency checks of the Standard Model.

Finally, it should be pointed out that the discussion presented
so far is strictly true only at Born level. At higher orders,
radiative corrections act to change these relations somewhat.
This is precisely what makes precision measurements interesting,
in that new physics in radiative loops will potentially yield
values for the observables in disagreement with Standard Model
expectations. On the other hand, `uninteresting' radiative
corrections, to the extent that they are not empirically
constrained, can enter in to obscure the above relations.
As will
be seen below, a particularly important example of this is the
scale dependence of $\alpha_{\rm em}$, which receives contributions
from intermediate-energy non-perturbative hadronic loops
and thus is not known to arbitrary precision.

\section{$Z^0$ Peak Facilities}

\begin{table}[t]
\caption{Sample Sizes (No. of Hadronic $Z^0$'s)
and Beam Polarization for the LEP and SLC
  Colliders  \label{tab:LEPSLC}}
\begin{center}
\begin{tabular}{|l|c|c|c|}
\hline
         &  LEP Sample     & SLC Sample       &                     \\
Year     &$\times 10^{-3}$ &$\times 10^{-3}$  & $<P_e>$ for the SLC \\
\hline
\hline
1988     &    0         &  0.5     & 0.0\%     \\
1989     &    0         &  0.3     & 0.0\%     \\
1990-91  & 1700         &  0.3     & 0.0\%     \\
1992     & 2800         &  10      & 22\%      \\
1993     & 2600         &  50      & 63\%      \\
1994     & 5800         &  70      & 78\%      \\
1995     & 2700         &  30      & 78\%      \\
1996     &    0         &  50      & 78\%      \\
1997-98  &    0         & 350      & 73\%      \\
\hline
{\bf Total}   & {\bf 15500}  & {\bf 550}  &     \\
\hline
\end{tabular}
\end{center}
\end{table}

Over the past decade, $Z^0$ peak running has been done at two
facilities: the LEP electron-positron storage ring collider at
CERN, and the SLC linear collider at SLAC. Table~\ref{tab:LEPSLC}
shows the relative sample sizes (summed over the four LEP detectors)
at the two facilities.

The SLC electron
beam polarization,
as discussed above, greatly increases the information content
per hadronic event for many precision measurements, making the
two facilities roughly comparable and nicely complementary.
In addition, the small luminous region of the SLC
(roughly 2 $\mu$m $\times$ 0.7 $\mu$m for the SLC, {\it vs.}
       250 $\mu$m $\times$ 5 $\mu$m for LEP), as well as the
smaller SLC interaction region beampipe radius (2.5 cm {\it vs.}
5.5 cm) give the SLC an intrinsic advantage for measurements
making use of precision tracking --
particularly those involving the production and decay of heavy flavor.

\section{$Z^0$ Pole Detectors}

\begin{figure}[t]
\epsfxsize=15pc 
\epsfbox{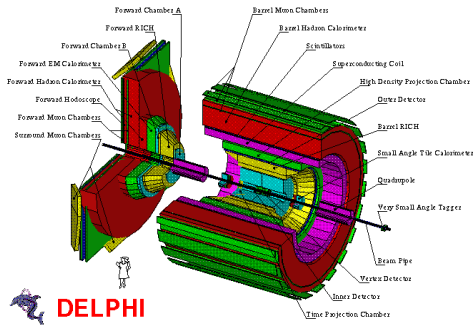} 
\caption{The DELPHI Detector at LEP  \label{fig:delphi}}
\end{figure}

The LEP facility is instrumented with four cylindrical geometry
detectors (ALEPH, DELPHI, L3, and OPAL), while the single
interaction region of the SLC was instrumented
with the upgraded MARK-II detector through 1989, and the SLD
detector thereafter. A schematic of the DELPHI detector is
shown in Figure~\ref{fig:delphi}. The primary components of this
detector, fairly typical of the six detectors that have run at the
$Z^0$, are as follows. Precise charged particle reconstruction is
done by a gaseous central tracker, with several precise space
points near the IP provided by a silicon $\mu$-strip vertex detector.
Calorimetry is mounted inside the magnet coil, outside of which is
placed coarse calorimetry and tracking to catch the hadronic tail
and identify muons. For the DELPHI and SLD detectors, a
Cerenkov ring-imaging detector resides between the tracking and calorimetry
to provide dedicated particle identification.
Forward instrumentation typically extends the
$|\cos \theta|$ coverage to 0.9 or better for tracking and 0.95 to 0.99
for calorimetry. Far-forward position 
sensitive electromagnetic calorimetry,
typically covering the angular region between 20 and 100 mrad, is used
to measure the interaction region luminosity via electrons from t-channel
Bhabha scattering.

Four advances in instrumentation technique which have
been primarily instigated by $Z^0$ pole physics are discussed in more
detail below.

\subsection{Cylindrical Geometry Silicon Tracking}

\begin{figure}[t]
\epsfxsize=13pc 
\epsfbox{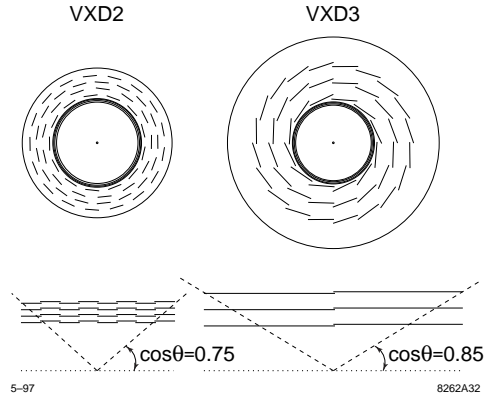} 
\caption{Schematic of the SLD CCD Pixel Tracking Detectors
                 \label{fig:CCD}}
\end{figure}

The first cylindrical geometry silicon $\mu$-strip detector was installed
in the MARK-II detector at the SLC in 1989.
By 1994 all of the $Z^0$ pole
detectors featured precision silicon tracking, achieving impact parameter
resolution at high momentum of better than 20 $\mu$m. The current state-of-the-art
silicon tracking device is the SLD pixel CCD detector, shown in two
different incarnations in Figure~\ref{fig:CCD}. The latest version (VXD3),
taking advantage of 8cm silicon wafer technology, comprises three layers
between 2.8 and 4.8 cm in $r$, providing 3 precise 3-d space points out to
$|\cos \theta|$ = 0.85. The CCD pixel dimension is $22 \times 22 \times 300$
$\mu$m, providing a single-hit resolution of approximately 4 $\mu$m. Including
the uncertainty in the beam spot location, VXD3 achieves an $r-\phi$
impact parameter resolution of 8 $\mu$m at high momentum, and 36 $\mu$m
at $p_{\perp} \sqrt{\sin \theta}$ = 1 GeV/c.

\subsection{Cerenkov Ring Imaging Detectors}

\begin{figure}[t]
\epsfxsize= 8pc 
\epsfbox{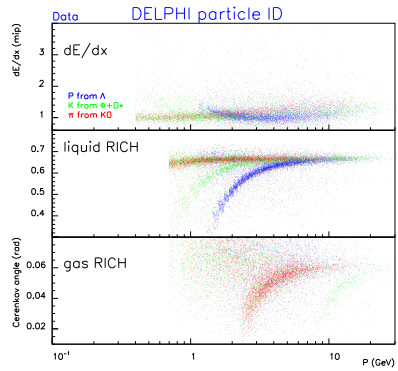} 
\caption{Performance of the DELPHI Particle ID System
 \label{fig:pid}}
\end{figure}

DELPHI and the SLD contain detectors which detect the photons emitted
by charged particles in the
well-defined Cerenkov cone as they pass through radiating materials.
Photons from a low-threshold, compact liquid radiator are converted
in a position-sensitive drift detector filled with a UV-sensitive gas.
Photons from a high-threshold gas radiator are imaged back into the
drift detector planes by curved mirrors mounted to the outer radius
of the Cerenkov detector. The radius of the reconstructed Cerenkov
photon ring is a measure of the angle of the Cerenkov cone, and thus
the velocity of the radiating particle, which, when combined with
momentum measured in the tracking system, provides a measure of
the particle's mass. Figure~\ref{fig:pid} shows the performance
of the liquid and gas Cerenkov systems, as well as the dE/dX
information from the gaseous tracker,
for the DELPHI detector. Taken together, these
systems provide reliable particle species separation over
much of the available kinematic range.

\subsection{Electron Beam Polarimetry}

The SLD electron beam polarization is analyzed via Compton scattering
from a polarized Nd:YAG laser beam. The magnet lattice just downstream
of the final focus serves to momentum analyze the scattered electrons,
beyond which the scattered electrons are detected in a position-sensitive
Cerenkov detector. The electron beam polarization modulates
the observed asymmetry in the Compton cross section between
aligned and anti-aligned laser and electron beam polarization, which is
measured in the Cerenkov detector as a function of scattered electron
energy to a relative precision of $\pm 0.7\%$. 

\subsection{Precise Luminosity Calibration}

The leading-order Bhabha acceptance at small angles is given by
\begin{equation}
\label{eq:bd2}
   \sigma_{e^+e^- \rightarrow e^+e^-} = {1040 nb-GeV^2 \over s}
 ({1 \over \theta_{min}^2} - {1 \over \theta_{max}^2}).
\end{equation}
Typical acceptance regions of between 20 and 100 mrad lead to
cross sections 3-10 times that of the $Z^0$ peak hadronic cross section.
The combination of precise, well-calibrated Si/W calorimeters and higher
order calculations of the Bhabha scattering cross-section has lead to
an overall relative error on the luminosity scale approaching $\pm 0.1\%$.

\section{$Z^0$ Lineshape Measurements}

\begin{figure}[t]
\epsfxsize=10pc 
\epsfbox{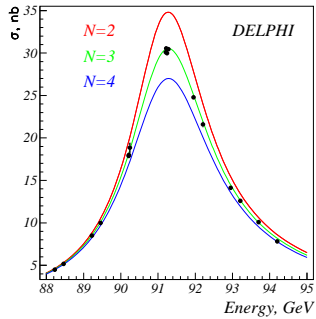} 
\caption{DELPHI Energy Scan Results (N is the number of light
$\nu$ species assumed). \label{fig:z0line}}
\end{figure}

\begin{table}[t]
\caption{Measured $Z^0$ Lineshape Parameters  \label{tab:z0par}}
\begin{center}
\begin{tabular}{|l|c|}
\hline
$M_Z$  & $91.1867 \pm 0.0021$ GeV/c$^2$ \\
\hline
$\Gamma_Z$   &  $2.4939 \pm 0.0024$ GeV/c$^2$   \\
\hline
$\sigma_{\rm had}^0$ &$41.491 \pm 0.058$ nb \\
\hline
\end{tabular}
\end{center}
\end{table}

The precise knowledge of the luminosity scale permits a correspondingly
precise measurement of the 
energy dependence of the $Z^0$ production cross-section,
from which accurate values of the lineshape parameters can be extracted.
Figure~\ref{fig:z0line} shows the results of such an energy scan, making
use of all available off-peak running, from the DELPHI collaboration. 
Table~\ref{tab:z0par}
shows the resulting lineshape parameters, averaged
for the four LEP experiments~\cite{data}.
The relative accuracy on the $Z^0$ mass
($2 \times 10^{-5}$) rivals that of the Fermi coupling constant $G_F$
($1 \times 10^{-5}$) and thus places a very tight constraint on the value
of the weak mixing angle via equation~(\ref{eq:bd1}).

\section{Forward-Backward Asymmetry Measurements}

In measuring the forward-backward asymmetries (polarized or
unpolarized) there
are two basic challenges to be met, both of which are relatively
straightforward for final state fermions $f = lepton$.
First, the $Z^0 \rightarrow f {\overline f}$ final state of interest must
be identified.
Secondly, the fermion direction must be determined -- usually by
finding the thrust axis, and then signing it according to the
best estimate of which hemisphere contains the fermion.

Calculating the mass of identified secondary vertices
can tag events with heavy final state quarks ($b,c$) as well as discriminate
between the $b$ and $c$ final states themselves.
$Z^0 \rightarrow b {\overline b}$
events are tagged in this way with an efficiency as high as 60\%, and
with a purity as high as 98\%. The fermion direction of such
`inclusively tagged' events can be determined by the net
momentum-weighted hemisphere track charge, the sign of
identified kaons from the
$b \rightarrow c \rightarrow s$ cascade, or simply the net charge
of the tracks forming the vertex.
Heavy quark events can be tagged by leptons from semileptonic decay,
or by the exclusive reconstruction of a heavy meson state, such as
$D^+ \rightarrow K^- \pi^+ \pi^+$. For these latter approaches, the
charge of the lepton or reconstructed state indicates the charge
of the underlying heavy quark.

\begin{figure}[t]
\epsfxsize=12pc 
\epsfbox{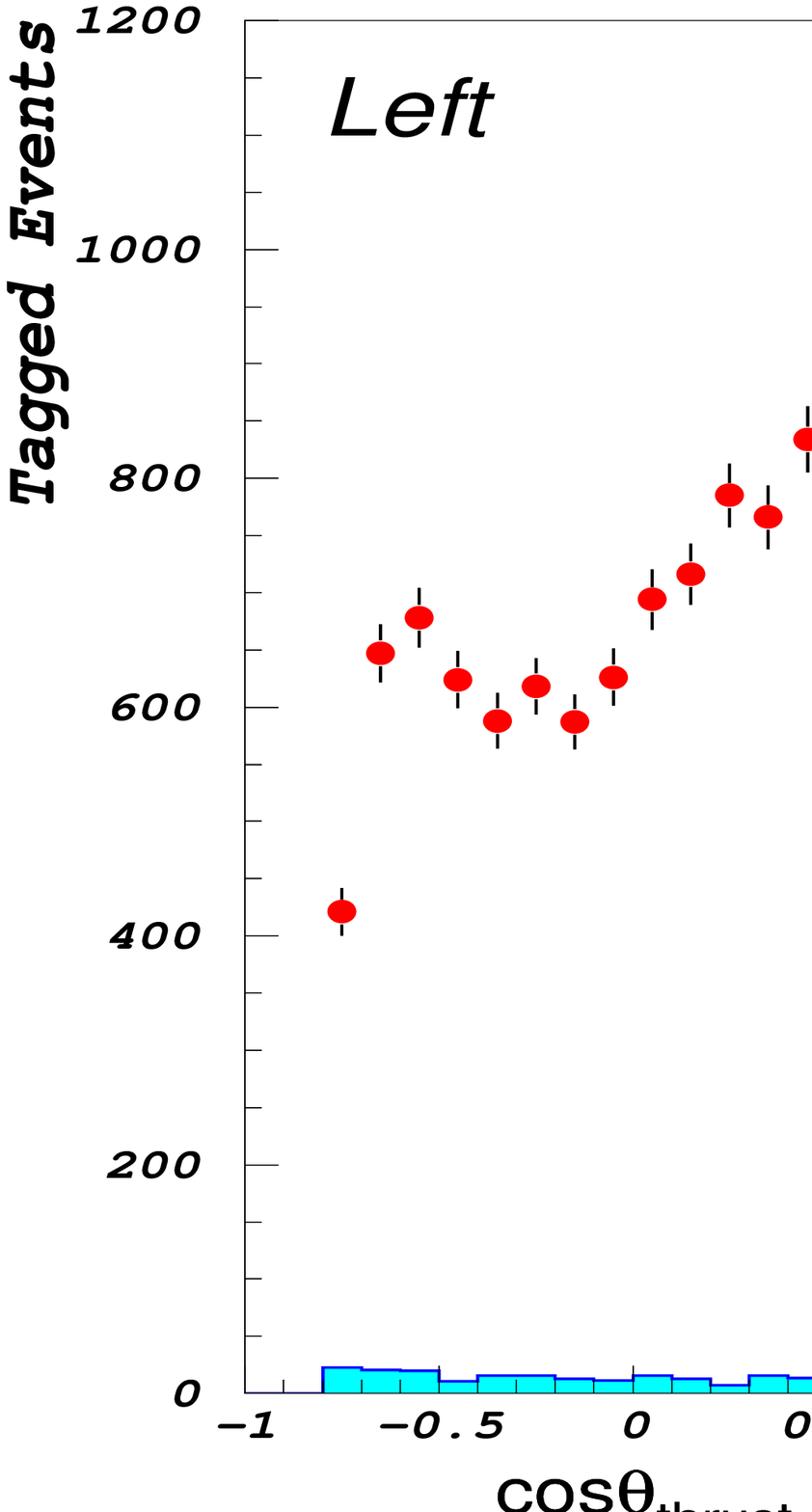} 
\caption{Distribution of the $b$ decay angle in SLD $Z \rightarrow
  b {\overline b}$ events, separately for left and right handed electron
beam.  \label{fig:abdist}}
\end{figure}

\begin{table}[t]
\caption{Forward-backward asymmetry measurements and associated
extracted electroweak parameters. For the unpolarized asymmetries,
the final state coupling parameters $A_f$ are derived under the
assumption that $A_e = 0.1489 \pm 0.0017$, which is the LEP/SLC
combined value for $A_{lepton}$.
\label{tab:afb}}
\begin{center}
\begin{tabular}{|l|c|c|}
\hline
                         Observable     &
                         Measured Value     &
                         Extracted EW Parameter     \\
\hline
\hline
$ {\rm A}_{FB}^l$  & $.01683 \pm .00096$         &
 $ {\sin^2 \theta_W} = .23117 \pm .00055    $  \\
$ {\rm A}_{FB}^b$  &  $.0991 \pm .0020$        &
 $ {\sin^2 \theta_W} = .23225 \pm .00037    $  \\
$ {\rm A}_{FB}^c$  &  $.0713 \pm .0043$        &
 $ {\sin^2 \theta_W} = .2321 \pm .0010      $  \\
$ {\rm A}_{FB}^{had}$ (${\rm Q}_{FB}$)&                          &
 $ {\sin^2 \theta_W} = .2321 \pm .0010      $  \\
$ {\tilde{\rm A}}_{FB}^l$  & $.1456 \pm .0063 $ &
 $ {\sin^2 \theta_W} = .2317 \pm .0008      $  \vspace*{-5mm}\\
                    &                             & \\
$ {\tilde{\rm A}}_{FB}^b$  & $.898 \pm .029$  &
 $ {A_b} =  .898 \pm .029                 $  \\
$ {\tilde{\rm A}}_{FB}^c$  & $.634 \pm .027$  &
 $ {A_c} =  .634 \pm .027                $  \\
$ {\rm A}_{FB}^b$  &        See caption          &
 $ {A_b} =  .887 \pm .021                   $  \\
$ {\rm A}_{FB}^c$  &        See caption          &
 $ {A_c} =  .637 \pm .039                   $  \\
\hline

\end{tabular}
\end{center}
\end{table}

Figure~\ref{fig:abdist} shows an example of the angular distribution
of the estimated $b$ quark direction, in this case for the
polarized electron beam at SLAC. Clear forward-backward
asymmetries are observed for both left- and right-handed beam, from
which the coupling parameter $A_b$ can be 
measured according to~(\ref{eq:dxs})
and~(\ref{eq:afbp}). Table~\ref{tab:afb}
shows the resulting electroweak
parameter results from the various forward-backward asymmetry
measurements~\cite{data}.

\section{Final State Polarization of $\tau$ Leptons}

An independent, direct measurement of the parity violation in the
$Z^0$-$\tau$ coupling can be made by measuring the asymmetry in
left- and right-polarized $\tau$ leptons in $Z^0 \rightarrow
\tau {\overline \tau}$ events, which is experimentally accessible
via the kinematic distributions of the $\tau$ decay products.
For example, a diagram of the decay $\tau \rightarrow
\nu_{\tau} \pi (K)$ similar to that of Figure~\ref{fig:angmom}
shows that the decay distribution in the angle $\theta^{\ast}$ 
of the $\pi$ (K) relative to the $\tau$ spin direction in the
$\tau$ rest frame is proportional to
$|{\rm d}^{1/2}_{-1/2,1/2}|^2 = \cos^2(\theta^{\ast}/2).$
Boosting against the direction of the $\tau$ spin (to produce
a right-handed $\tau$) results in a distribution
\begin{equation}
\label{eq:bd3}
   {{\rm d}\Gamma^{\tau}_{RH} \over {\rm d}x} \propto  x
\end{equation}
where $x = E_{\pi,K}/E_{beam}$ is the fractional energy of
the pseudoscalar decay product. Similarly, for a boost
in the direction of the $\tau$ spin (to produce a LH $\tau$),
the decay angular distribution is proportional to $1-x$, and
so the relative fraction of RH and LH polarized $\tau$'s can
be extracted from the $x$ distribution of 1-prong $\tau$ decays.

The relation between this $\tau$ polarization asymmetry and
the electroweak coupling parameters is given by
\begin{equation}
\label{eqq:bd4}
  P_{\tau}(z) = {\sigma_L^{\tau}(z) - \sigma^{\tau}_R(z) \over
                 \sigma_L^{\tau}(z) + \sigma^{\tau}_R(z) }  =
{{A_{\tau}}(1+z^2) + 2{A_e}z \over
 (1+z^2) + 2{A_{\tau} A_e}z}
\end{equation}
and so a measurement of $P_{\tau}(z)$ yields independent values for
$A_{\tau}$ and $A_e$. Table~\ref{tab:tpol} shows the $\tau$
polarization results, averaging over all analyzable $\tau$ decay
modes for the four LEP experiments~\cite{data}.

\begin{table}[t]
\caption{$\tau$ Polarization Asymmetry Results
\label{tab:tpol}}
\begin{center}
\begin{tabular}{|l|c|c|}
\hline
                         Parameter      &
                         Measured Value     &
                         Extracted $\sin^2\theta_W$  \\
\hline
\hline
   $A_{\tau}$    &  0.1431 $\pm$ 0.0045   & 0.23202 $\pm$ 0.00057 \\
   $A_e     $    &  0.1479 $\pm$ 0.0051   & 0.23141 $\pm$ 0.00065 \\
\hline
\end{tabular}
\end{center}
\end{table}

\section{The Left-Right Asymmetry $A_{LR}$}

The left-right asymmetry $A_{LR}$ is measured by simply comparing the
number of hadronic $Z^0$ decays produced with left- and right-handed
electron beam:
\begin{equation}
\label{eq:bd5}
  A_{LR} = {1 \over |P_e|} {N_{Z^0}^L - N_{Z^0}^R \over
            N_{Z^0}^L + N_{Z^0}^R}.
\end{equation}
Corrections due to differences in left- and right-handed beam
luminosity, energy, polarization, etc., are small and well
understood. With a sample of 550,000 hadronic $Z^0$ decays
produced with $\langle |P_e| \rangle \simeq 72 \%$, the SLD
experiment finds~\cite{data}
\begin{equation}
\label{eq:bd6}
\begin{array}{rcl}
   A_e = A_{LR} = 0.1511 \pm 0.0024        \\[4pt]
   \sin^2\theta_W = 0.23100 \pm 0.00031.
\end{array}
\end{equation}
In addition to being the single most precise determination of the
weak mixing angle, the systematics of this measurement are
almost entirely independent of that of other techniques discussed
above.

\section{Combined Weak Mixing Angle Results and Fits}

\begin{figure}[t]
\epsfxsize=10pc 
\epsfbox{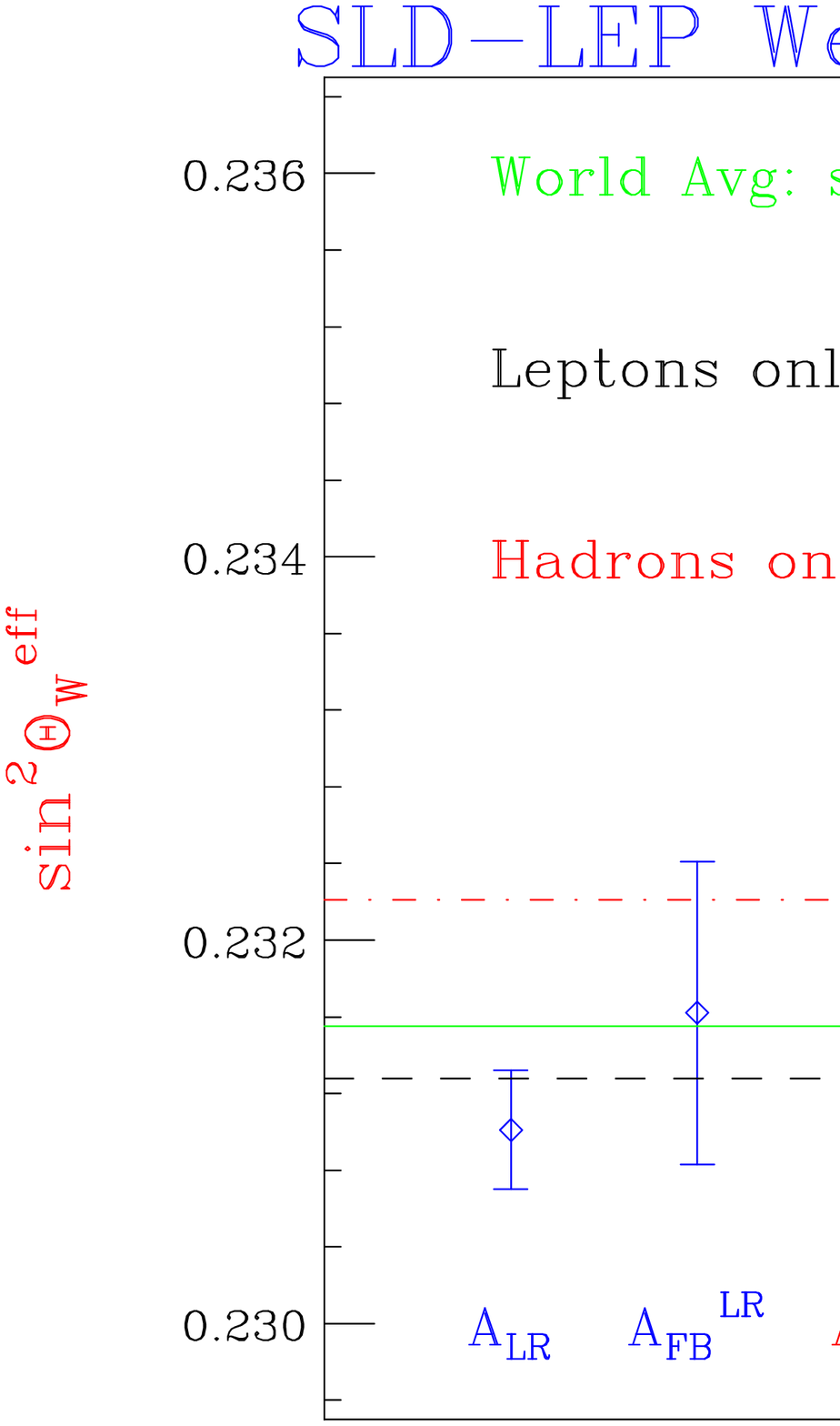} 
\caption{Results for $\sin^2 \theta_W$ By Method
  \label{fig:s2meth}}
\end{figure}

\begin{figure}[t]
\epsfxsize=12pc 
\epsfbox{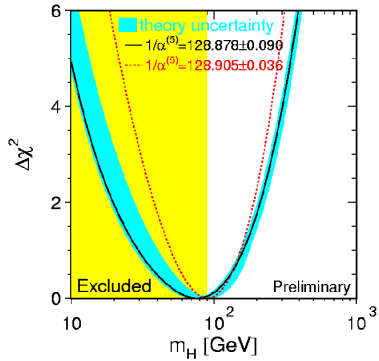} 
\caption{Fit of the SM Higgs Boson mass to electroweak precision data.
The solid curve uses as input the value of
$\alpha_{\rm em}(Q^2 = M_Z^2)$ of Ref. [3], while the dashed curve
uses that of Ref. [4].
\label{fig:smfit}}
\end{figure}

Figure~\ref{fig:s2meth} shows the $\sin^2 \theta_W$ results
separately for each of the above approaches~\cite{data}. While the
hypothesis that all of these measurements are consistent with
a single mean value is reasonably well supported
($\chi^2 = 8.0$ for 7 d.o.f), there is
a mild (2.2 $\sigma$) discrepancy between the values measured
with leptonic final states and those measured with hadronic
final states, the latter of which are dominated by the unpolarized
${\rm A}_{FB}^b $ measurement. Ignoring this issue for
the moment,
Figure~\ref{fig:smfit} shows a fit for the Higgs Boson mass, assuming
the Higgs contributes to radiative loops as prescribed by the Standard
Model.
At 95\% confidence,
the Higgs mass is restricted to be
below about 270 GeV/c$^2$. 

\begin{table}[t]
\caption{Partial Width Measurements
\label{tab:pwid}}
\begin{center}
\begin{tabular}{|l|c|c|}
\hline
                         Parameter      &
                         Measured Value     &
                         SM Expectation              \\
\hline
\hline
   $R_b     $    &  0.2168 $\pm$ 0.0007   & 0.2155 $\pm$ 0.0003($m_t$)\\
   $R_c     $    &  0.1694 $\pm$ 0.0038   & 0.1723                \\
\hline
\end{tabular}
\end{center}
\end{table}

\section{Partial Width Measurements}

Measurements of $Z^0$ decay partial widths~(\ref{eq:pwd})
are particularly sensitive to vertex corrections. Due to the
cancelation of both experimental and theoretical uncertainties
(luminosity, fiducial volume, effects of gluon radiation, etc.), the
partial widths are most cleanly constrained via hadronic branching
fractions, e.g. $R_b = \Gamma_{Z^0 \rightarrow b{\overline b}}/
    \Gamma_{Z^0 \rightarrow {\rm had}}$.
In addition, with the pure and efficient tags provided by precise
silicon tracking, it is possible to constrain the final-state
tagging efficiency by comparing the single- {\it vs.} double-hemisphere
tagging rate. The resulting precise constraints on the $b$ and $c$
quark partial widths are shown in Table~\ref{tab:pwid}.

\section{Fit for $Z^0 - b$ Vertex Parameters}

\begin{figure}[t]
\epsfxsize=18pc 
\epsfbox{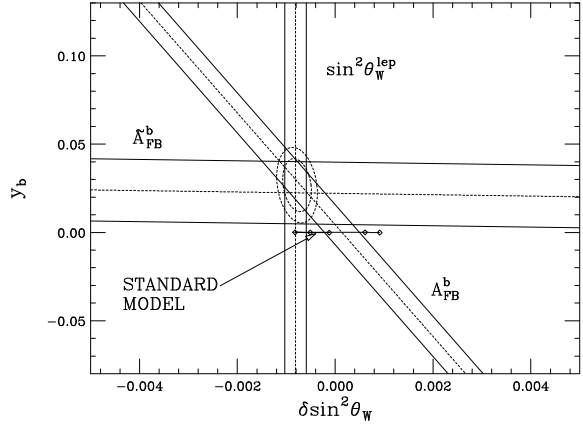} 
\caption{Fit for $Z^0-b$ vertex parameters, projected into the
$\delta \sin^2 \theta_W$--$y_b$ (parity-violation) plane. The
Standard Model range includes $169 < m_t < 179$ GeV/c$^2$, and
$100 < m_H < 1000$ GeV/c$^2$.
\label{fig:tgr}}
\end{figure}

As discussed in Ref. [5],
it is interesting to decouple potential
anomalous $Z^0 - b$ vertex effects from the vacuum polarization corrections
to the weak mixing angle.
The explicit $\sin^2 \theta_W$ dependence of the
$Z^0 - b$ coupling is removed by writing
\begin{equation}
\label{eq:bd7}
\begin{array}{rcl}
  {\delta s} = [\sin^2 \theta_W]_{meas} - [\sin^2 \theta_W]_{pred}
 \\ [6pt]
  {1 \over 3} {\delta s} + {\delta g_L^b} =
[g_L^b]_{meas} - [g_L^b]_{pred}    \\ [6pt]
  {1 \over 3} {\delta s} + {\delta g_R^b} =
[g_R^b]_{meas} - [g_R^b]_{pred}
\end{array}
\end{equation}
Then, the linear combinations
\begin{equation}
\label{eq:bd8}
\begin{array}{rcl}
  {x_b} = \cos \phi  {\delta g_L^b}
- \sin \phi  {\delta g_R^b} \simeq -{1 \over 4} \delta R_b \\ [6pt]
  {y_b} = \sin \phi  {\delta g_L^b}
+ \cos \phi {\delta g_R^b} \simeq -{3 \over 5} \delta A_b,
\end{array}
\end{equation}
with $\phi = \tan^{-1} |{g_R^b} / {g_L^b} |,$ are the deviations
from the SM in the $Z^0-b$ vertex partial width ($x_b$) and parity
violation ($y_b$) strengths. 
These parameters are constrained by the precision measurements of
$\sin^2 \theta_W^{\rm lepton}$, $R_b$, $R_c$,
${\tilde {\rm A}}_{FB}^b$, ${\rm A}_{FB}^b$,
$R_Z = \Gamma_{\rm had}/\Gamma_{\mu^+ \mu^-}$, and $\sigma^0_{\rm had}$.
Of particular interest, given the
high value of $\sin^2 \theta_W$ from ${\rm A}_{FB}^b$, is the projection
of this fit into the ($\delta s$,$y_b$) plane, shown in 
Figure~\ref{fig:tgr}. It is seen that the high value of
$\sin^2 \theta_W$ from $A_{FB}$ (see Table~\ref{tab:afb} or
Figure~\ref{fig:s2meth}) may possibly be due to an anomalous
$Z^0-b$ vertex coupling parameter, a hypothesis which is
supported by the direct measurement provided by
${\tilde {\rm A}}_{FB}^b$. The overall discrepancy in the strength
of $Z^0-b$ coupling parity violation with respect to the SM
is just over $2.5 \sigma$.

\section{Summary and Outlook}

With the exception of relatively minor updates to existing
results, the picture provided by $Z^0$ resonance precision measurements
is essentially complete.
The totality of accumulated data
has allowed, with a number of systematically independent
approaches, the test of the internal consistency of the
Standard Model to the unprecedented precision of
$\delta \sin^2 \theta_W < \pm 0.0002$ -- roughly 300 times
more accurate than the $\pm \sim 0.06$ available from
$\nu$ DIS in 1988. While this confirmation of the predictions
of the Standard Model is remarkable, there are nonetheless
a couple of issues to be noted. A mild discrepancy between
the weak mixing angle determined via purely leptonic couplings
relative to that made with hadronic couplings could possibly be
due to anomalous $Z^0-b$ vertex effects. This hypothesis is
supported, albeit with somewhat inadequate experimental precision,
by the SLD direct measurements of the $Z^0-b$ vertex via the
polarized forward-backward asymmetry
${\tilde {\rm A}}_{FB}^b$. Secondly, the combination of all
electroweak data, assuming only Standard Model contributions
to radiative loops, is inconsistent with a heavy Higgs Boson.
Should the minimal Standard Model be correct,
the Higgs should easily be
seen at the LHC, or perhaps earlier in the imminent RUN II of
the Fermilab Tevatron.

\section*{Acknowledgment}
This work was supported in part by the Department of Energy,
Grant \#DE-FG03-92ER40689.

\end{document}